\documentclass[article,pdftex,oneauthor,accept,preprints,entropy]{mdpi}
\usepackage[utf8]{inputenc}
\usepackage{float}
\usepackage{textcomp}
\usepackage{graphicx}
\ifx\hypersetup\undefined
  \AtBeginDocument{%
    \hypersetup{unicode=true}
  }
\else
  \hypersetup{unicode=true}
\fi

\makeatletter

\Title{Comparison of Step Samplers for Nested Sampling}

\TitleCitation{Comparison of Step Samplers for Nested Sampling}

\Author{Johannes Buchner$^{1,*}$\orcidA{}}

\AuthorNames{Johannes Buchner}

\AuthorCitation{Buchner, J. }

\address{$^{1}$\quad{}Max Planck Institute for Extraterrestrial Physics,
Giessenbachstrasse, 85741 Garching, Germany}

\corres{Correspondence: johannes.buchner.acad@gmx.com}

\abstract{Bayesian inference with nested sampling requires a likelihood-restricted
prior sampling method, which draws samples from the prior distribution
that exceed a likelihood threshold. For high-dimensional problems,
Markov Chain Monte Carlo derivatives have been proposed. We numerically
study ten algorithms based on slice sampling, hit-and-run and differential
evolution algorithms in ellipsoidal, non-ellipsoidal and non-convex
problems from 2 to 100 dimensions. Mixing capabilities are evaluated
with the nested sampling shrinkage test. This makes our results valid
independent of how heavy-tailed the posteriors are. Given the same
number of steps, slice sampling is outperformed by hit-and-run and
whitened slice sampling, while whitened hit-and-run does not provide
as good results. Proposing along differential vectors of live point
pairs also leads to the highest efficiencies, and appears promising
for multi-modal problems. The tested proposals are implemented in
the \href{https://johannesbuchner.github.io/UltraNest/}{UltraNest}
nested sampling package, enabling efficient low and high-dimensional
inference of a large class of practical inference problems relevant
to astronomy, cosmology, particle physics and astronomy. }


\providecommand{\tabularnewline}{\\}

\firstpage{1} 
 
\pubvolume{1}
\issuenum{1}
\articlenumber{0}
\pubyear{2022}
\copyrightyear{2022}
\datereceived{} 
\dateaccepted{} 
\datepublished{} 
\hreflink{https://doi.org/} 

\makeatother

\begin{document}
\maketitle

\section{Introduction}

The nested sampling Monte Carlo algorithm (proposed by \citep{Skilling2004}
and recently reviewed in \citep{Ashton2022} and \citep{Buchner2021})
enables Bayesian inference by estimating the posterior and its integral.
A population of $K$ live points are sampled randomly from the prior.
Then, the lowest likelihood point is discarded, and a new live point
is sampled from the prior under the constraint that its likelihood
must by higher than the discarded point. This is iteratively repeated
in nested sampling. The likelihood-restricted prior sampling (LRPS),
if unbiased, causes the likelihoods of discarded points to have interesting
properties. In particular, the fraction of prior mass below the likelihood
threshold is approximately $1/K$. The recursive estimation of prior
mass discarded, together with the sampled likelihood, allows nested
sampling to perform the integration.

The first LRPS algorithms proposed were adaptations of Markov Chain
Monte Carlo (MCMC) variants. For simplicity, here we assume that the
parameters have been parameterized so that the prior probability density
is uniform. Then, in a Gaussian Metropolis MCMC started from a randomly
chosen live point a random nearby point would be proposed. If the
likelihood there exceeds the current threshold, it replaces the starting
point for the next iteration. This procedure is repeated $N_{\mathrm{steps}}$
times, after which the final point is considered a sufficiently independent
prior sample. Since the value of the likelihood is not considered
further, this pursues a purely geometric random walk. The class of
such step sampler algorithms \citep{Buchner2014stats} includes slice
sampling \citep{neal2003}, first proposed for LRPS by \citep{jasa2005using}.
In section §\ref{subsec:LRPS-Methods-considered} we describe ten
members of a larger family of algorithms which includes slice sampling.
These algorithms are popular because no additional parameters that
influence its outcome need to be chosen, making applications to a
wide variety of problems robust. When the dimensionality $d$ of the
inference problem is high ($d\gg20$), i.e., the fitted model has
many parameters, such step samplers tend to outperform rejection sampling-based
algorithms (see \citep{Buchner2021} for a survey).

Given a inference problem with $d$ parameters, choosing a LRPS method
and its parameters is not trivial. Detecting biased LRPS is a question
of on-going research (e.g., \citep{Buchner2014stats,Higson2019,Fowlie2020}),
and we give our definition of acceptable LRPS quality in §\ref{subsec:Quantifying-the-goal}.
Our procedure for calibrating $N_{\mathrm{steps}}$ is described in
§\ref{subsec:Calibrating-Nsteps}. The test problems are listed in
§\ref{subsec:Geometries-considered} and the LRPS methods described
in §\ref{subsec:LRPS-Methods-considered}. The evaluation results
of the mixing behaviour of the various LRPS methods is presented in
§\ref{subsec:Calibration-results}, and discussed in §\ref{sec:Discussion}.

\section{Materials and Methods}

\subsection{Distinguishing good and bad LRPS performance \label{subsec:Quantifying-the-goal}}

How can we judge that one method performs better than another, or
even acceptably well? We could run NS on example inference problems
and analyse the produced posterior and integral. While this may be
very realistic, the results would be limited to the geometry and likelihood
slopes of that particular inference problem, leaving generalization
to different data or models unclear. It is also difficult to define
an objective criterion on whether the integral and posterior approximation
is acceptable, in addition to the requirement that the true result
is available for comparison. Alternative to such a ``in vivo'' test,
we could explore the geometric mixing behaviour in isolation (``in
vitro''). For example, from initial samples concentrated in a small
ball, we could measure the time until samples have converge to true
geometry. Here, the information gain could measure the quality, but
thresholds are also unclear. Additionally, in NS the situation is
not static, rather a sequence of similar geometries is explored with
ever-shrinking size. To address the shortcomings listed above, we
instead adopt a verification test built for NS that is already familiar.

The shrinkage test was presented by \citep{Buchner2014stats}. In
problems where the volume enclosed at a likelihood threshold, $V({\cal L})$,
can be computed, the shrinkage test analyses the volume ratio distribution
$V_{i+1}/V_{i}$ of a sequence of discarded samples produced by NS
with a LRPS. This measures the critical issue of interest, namely,
whether the shrinkage is passed to NS correctly (following the expected
beta distribution) by an unbiased LRPS.

The shrinkage test can flag problematic LRPS methods. We collect shrinkage
samples for $25\times K$ LRPS iterations, with $K=400$ live points.
Given these samples from $10,000$ iterations, configurations are
rejected if the shrinkage KS test gives a p-value below 0.01. We consider
a configuration the combination of a LRPS method together with its
hyperparameters. Additionally, if any iteration had no movement (live
point is stuck), the configuration is also rejected. The LRPS is started
with live points drawn perfectly from the geometry, and the first
$3\times K$ iterations are not included in sample collection to allow
the LRPS method to warm up and calibrate.

On real inference problems, troublesome situations may occur at some
arbitrary iteration and persist for some arbitrary length, both of
which modulate the impact on evidence and posterior estimation. The
shrinkage test is independent of this and thus provides conservative
results. With NS applications typically requiring several tens of
thousands iterations, our detections of deviations provide lower limits.
Importantly, the shrinkage test results are independent of the likelihood
slope and are therefore generalizable. Only the geometry of the likelihood
contours are important, regardless of how informative the posterior
is, whether there are phase transitions, etc. While the requirement
that $V({\cal L})$ can be computed analytically may seem restrictive,
this is possible for difficult geometries (§\ref{subsec:Geometries-considered}),
including ellipsoidal, multi-ellipsoidal, convex and non-convex problems.
We seek LRPS methods that behave robustly across such situations.

\subsection{Geometries considered\label{subsec:Geometries-considered}}

Likelihood contours can be classified into convex or non-convex, depending
on whether the linear interpolation between two points within the
restricted prior is also inside. A sub-class of the convex geometries
are ellipsoidal contours, arising from Gaussian or other ellipsoidal
posterior distributions. This classification can also be considered
locally. Different information gain per parameter influences how the
contour changes shape. 

Our goal here is to span the space of geometries across dimensionality,
while simultaneously avoiding excessive computational cost. Therefore,
we sparsely sample the problem space of geometries with six setups.
This includes (1) as a ellipsoidal problem a correlated Gaussian with
covariance $\Sigma_{ij}=0.95$ and a unit diagonal, in 16 and 100
dimensions, (2) as a difficult convex problem the hyperpyramid with
log-likelihood $\log{\cal L}=\max_{i}\{|x|\}$ in four and 16 dimensions,
and (3) as a non-convex problem the Gaussian shell problem \citep{Feroz2008}
$\log{\cal L}=-((||x-0.5||^{2}-0.4^{2})/0.004)^{2}$ in two and eight
dimensions. The former two are self-similar across iterations, while
the last is not, with the shell thickness becoming thinner with increased
iteration. To avoid entering extreme situations infrequent in real
applications, short runs of length $3000$ ($d=2$) and $6000$ iterations
($d=8$) were performed with different seeds, and repeated until the
desired number of samples were collected.

The scope sampled by these geometries is limited, and does not capture
all real-world situations of interest. The scope of this work is to
develop a methodology to evaluate LRPS behaviour and to discard configurations
that are already showing problematic behaviour in these simple geometries.
However, there are some relevancies. For example, posteriors often
are Gaussians away from the borders, which is approximated by (1).
Configuration (2) approximates when the data impose upper limits on
each parameter independently. Non-linear banana degeneracies are a
sub-set of (3). Multi-modal problems are left for future work. However,
we can extrapolate in limited way: Methods that only consider local
instead of global live point correlations may be unimpacted by inter-mode
structure, and thus our results may hold for these. To emphasize,
some reasonable situations are considered here in-vitro to place some
limits on in-vivo behaviour, but we do not aim to consider the most
difficult case conceivable.

\subsection{Calibrating the number of steps\label{subsec:Calibrating-Nsteps}}

To find the $N_{\mathrm{steps}}$ required to sample the geometry
correctly, a run as described above with $N_{\mathrm{steps}}=1$ is
performed. Upon rejection, $N_{\mathrm{steps}}$ is doubled until
success. Since we want a lower limit $N_{steps}(d)$ usable by practitioners,
and because this should be a monotonically increasing function, an
improvement for computational efficiency can be made: If the test
problems are ordered by $d$, then the next problem does not need
to be tested on $N_{\mathrm{steps}}$ that already failed a previous
problem. So we run the problems in order of $d$ and initialize the
search for $N_{\mathrm{steps}}$ at the value that was acceptable
for the previous problem. This automatically finds a conservative
$N_{steps}(d)$ curve across the different problems.

\subsection{LRPS methods considered\label{subsec:LRPS-Methods-considered}}

Multiple algorithms are considered. Within a single step, they have
similar behaviour, which we describe first. From a starting point
$x$, a direction $v$ is chosen. To sample a point on the slice $y(t)=x+t\times v$,
with $t$ uniformly sampled, bounds on $t$ first need to be found.
Given a guess length $L$, initially $L=1$, the point at $t=L$,
$t=2L$ etc. is tested until the point lies outside the restricted
prior, $t=2^{i}L$. The same is repeated for negative $t$ values.
Then, $t$ can be uniformly sampled between $t_{-}$ and $t_{+}$.
If the point $y(t)$ is rejected, subsequent tries can set $t_{-}=t$
(if $t$ is negative) or $t_{+}=t$ (if $t$ is positive). This is
akin to bisection and assumes convexity \citep{neal2003}. When the
$y(t)$ point is accepted, the sampling is restarted from there. This
is done $N_{\mathrm{steps}}$ times. The number of model evaluations
per NS iteration is therefore $N_{\mathrm{steps}}\times(i_{+}+i_{-}+j)$,
where the $i$ stepping out iterations and $j$ bisection steps, are
subject to the constraint behaviour and the direction proposal. The
guess length $L$ is increased by $10\%$ if $i_{+}+i_{-}>0$, or
decreased by 10\% otherwise.

We consider ten algorithms, which differ in their direction proposal.
The algorithm variants are:
\begin{enumerate}
\item cube-slice: In slice sampling, one parameter is randomly chosen as
the slice axis \citep{neal2003}.
\item region-slice: Whitened slice sampling, randomly choosing a principal
axis based on the estimated covariance matrix.
\item region-seq-slice: Whitened slice sampling, iteratively choosing a
principal axis \citep{Speagle2020}. 
\item cube-harm: Hit-and-run sampling choses a randomly chosen direction
from the unit sphere (\citep{doi:10.1137/1116083}; combined with
slice bisection: \citep{Kiatsupaibul2011}).
\item region-harm: Whitened hit-and-run sampling by randomly choosing a
ball direction using the sample covariance.
\item cube-ortho-harm and 
\item region-ortho-harm are the same as cube-harm and region-harm, but prepare
a sequence of $d$ proposals that are made orthogonal by Gram-Schmidt,
then use them in order. This is similar to the PolyChord \citep{Handley2015a}
implementation.
\item de-harm: Differential evolution (randomly choose pair of live points
and use their differential vector) \citep{ter2006markov}
\item de1: same as de-harm, but set the difference in all but one, randomly
chosen dimension to zero. This is very similar to cube-slice, but
scales the direction using the distance between points.
\item de-mix: randomly choose between de-harm and region-slice at each step
with equal probability.
\end{enumerate}
The phase from proposing a slice until acceptance of a point on that
slice is considered a single slice sampling step for counting to $N_{\mathrm{steps}}$.
This is also the case for methods where subsequent proposals are prepared
together (e.g., region-seq-slice, cube-ortho-harm, region-ortho-harm).
Some algorithms take advantage of the sample covariance matrix estimated
from the live points, and its principle axis, to whiten the problem
space. We estimate the covariance matrix every $K/5$ nested sampling
iterations, i.e., when the space has shrunk to $(1-1/K)^{K/5}\approx80\%$. 

Detailed balance is satisfied by the cube-{*} and de-{*} methods.
When whitening is computed from the live points, detailed balance
is not retained. Since detailed balance is a sufficient but not a
necessary condition of the Metropolis proposal, this may or may not
lead to convergence issues. The shrinkage test should indicate this.

\section{Results}

\subsection{Calibration results}

\label{subsec:Calibration-results}

\begin{figure}[H]
\centering{}\includegraphics[width=0.8\columnwidth]{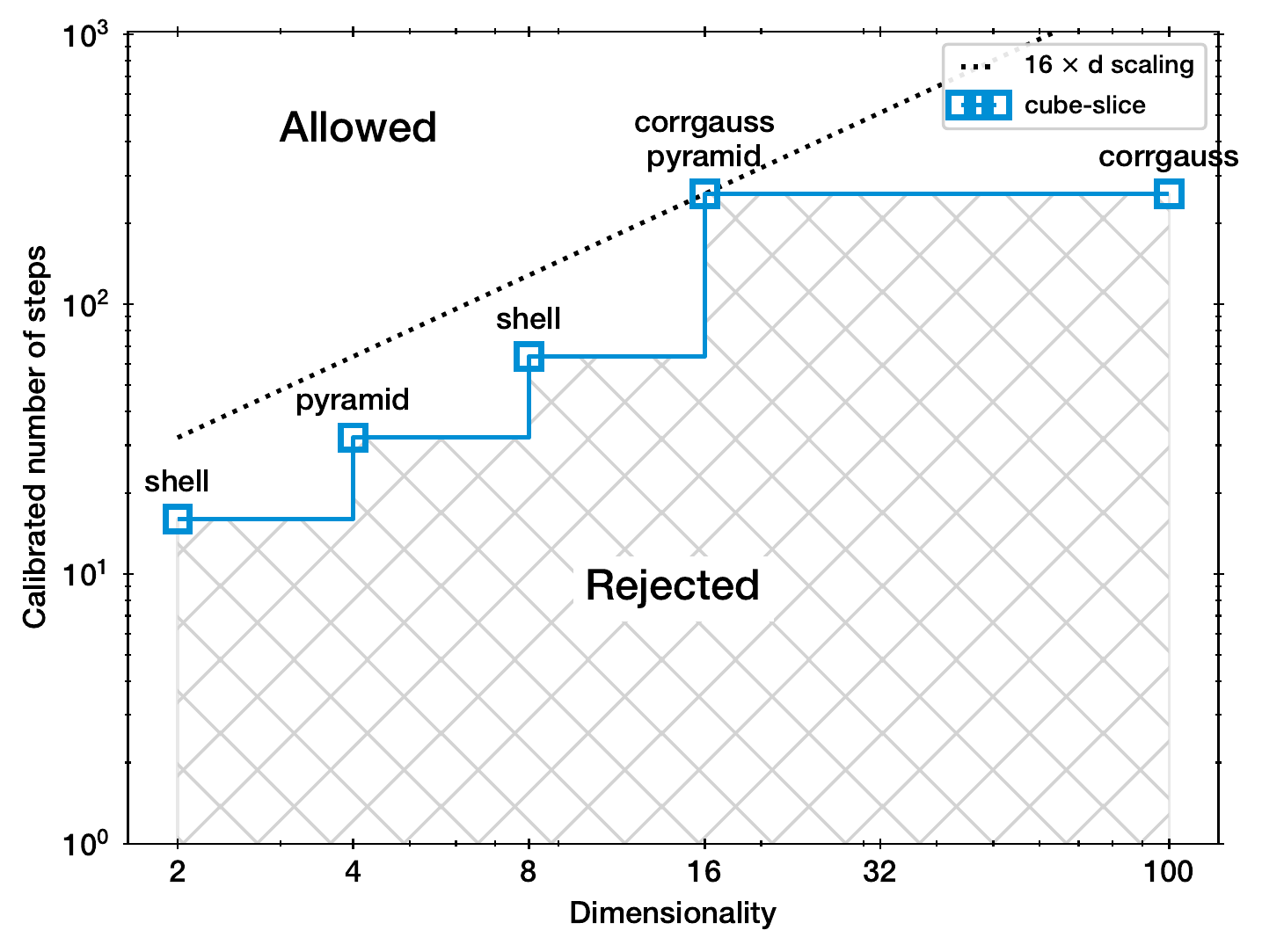}\caption{Calibration for the slice sampler (cube-slice).\label{sliceresult}}
\end{figure}

To start, Figure~\ref{sliceresult} presents the calibration results
for the cube-slice method. The $N_{\mathrm{steps}}$ needed for each
problem is shown by the blue curve. In this case, the $N_{\mathrm{steps}}$
needed to be doubled or quadrupled for all but the last increase in
dimensionality. Configurations to the bottom right of this curve are
rejected, because biases in the shrinkage distributions were detected.
Configurations in the upper left including the blue squares are still
allowed, but may be problematic in geometries other than the ones
considered here. The dotted black line shows the minimal scaling law
that avoids the rejected configuration space.

\begin{figure}[H]
\includegraphics[width=1\columnwidth]{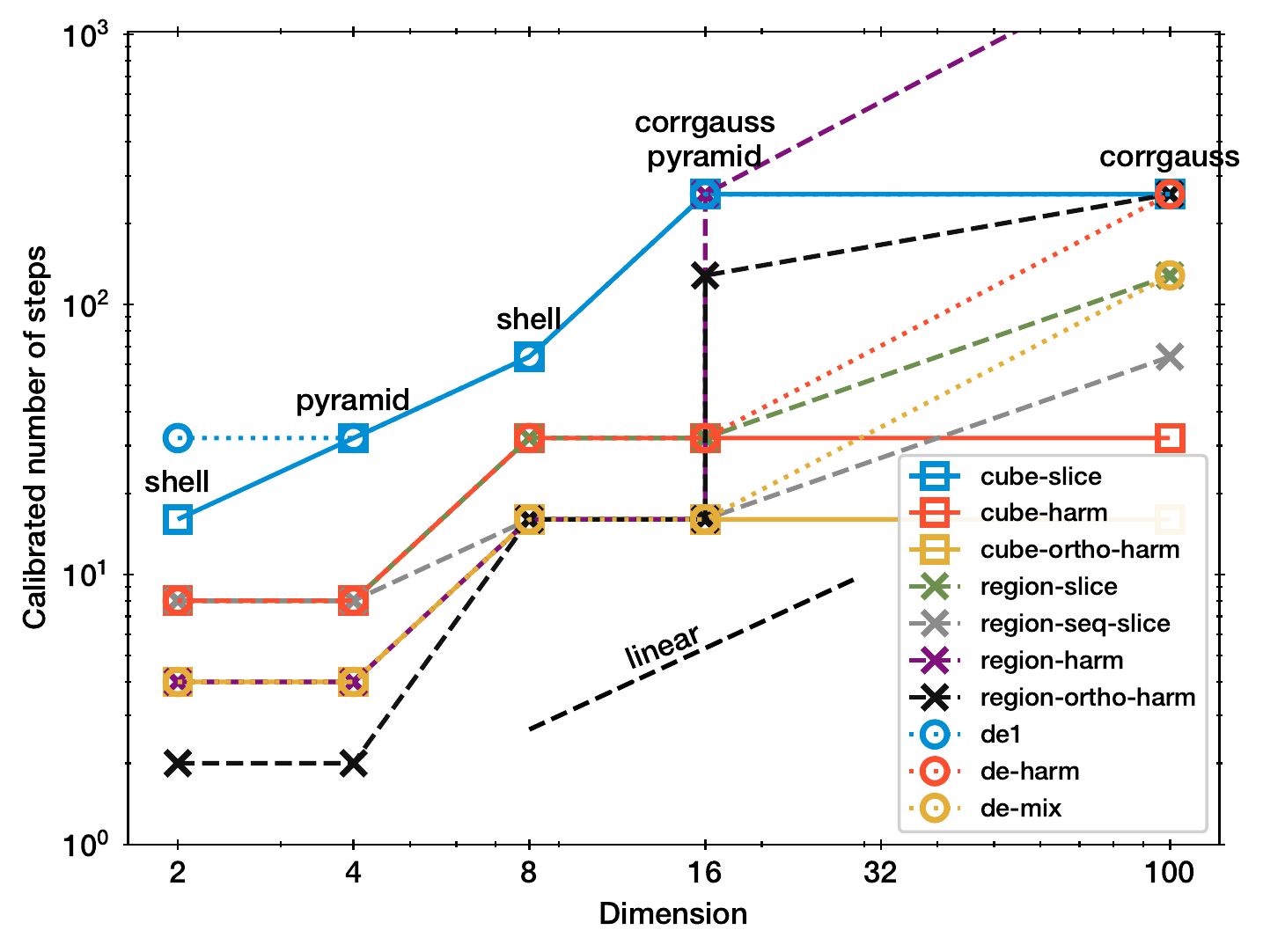}\caption{Calibration of all samplers.\label{sliceresult-2}}
\end{figure}

Next, Figure~\ref{sliceresult-2} present the same calibration, but
now for all step samplers. In general, all curves show an approximately
linear increase with $d$, although with different normalisations.

de1 performs the same as cube-slice (blue circles with dotted curve
and blue rectangles with solid curve). This is expected, because the
methods are identical except for the proposed length step. It suggests
our experimental setup gives reproducible results, even though the
KS threshold has some randomness.

\begin{table}[H]
\caption{Minimum scaling factor for a linear calibration $N_{\mathrm{steps}}=k\times d$.
With the efficiency $\epsilon$ defined as number of model evaluations
per iteration, the last column reports the lowest observed dimension-corrected
efficiency $\epsilon\times d$. \label{tab1}}

\begin{tabular*}{1\textwidth}{@{\extracolsep{\fill}}>{\centering}p{0.33\textwidth}>{\centering}p{0.33\textwidth}>{\centering}p{0.33\textwidth}}
\toprule 
\textbf{Sampler} & \textbf{Factor $k$} & \textbf{Efficiency \%}\tabularnewline
\midrule 
cube-slice & 16 & 0.32 \tabularnewline
cube-harm & 4 & 1.19 \tabularnewline
cube-ortho-harm & 2 & 2.32 \tabularnewline
region-slice & 4 & 1.16 \tabularnewline
region-seq-slice & 4 & 1.69 \tabularnewline
region-harm & >16 & --  \tabularnewline
region-ortho-harm & 8 & 1.42 \tabularnewline
de1 & 16 & 0.43 \tabularnewline
de-harm & 4 & 1.15 \tabularnewline
de-mix & 2 & 2.33 \tabularnewline

\bottomrule
\end{tabular*}

\end{table}

For the 100d problem, region-harm did not converge even for $N_{\mathrm{steps}}>1024$,
and is set to a arbitrary high value in this plot. The whitened hit-and-run
methods (region-harm, region-ortho-harm) need the most steps, together
with the de1. Only cube-harm needs few steps even in high-d. The calibrated
scaling factors for each method are listed in Table~\ref{tab1}.

Now that all methods are calibrated to perform correctly, we can compare
their efficiency in a fair way. This is not the same as the number
of steps, because the number of stepping-out iterations may differ.

\subsection{Computational cost scaling with dimension}

\begin{figure}[H]
\includegraphics[width=1\columnwidth]{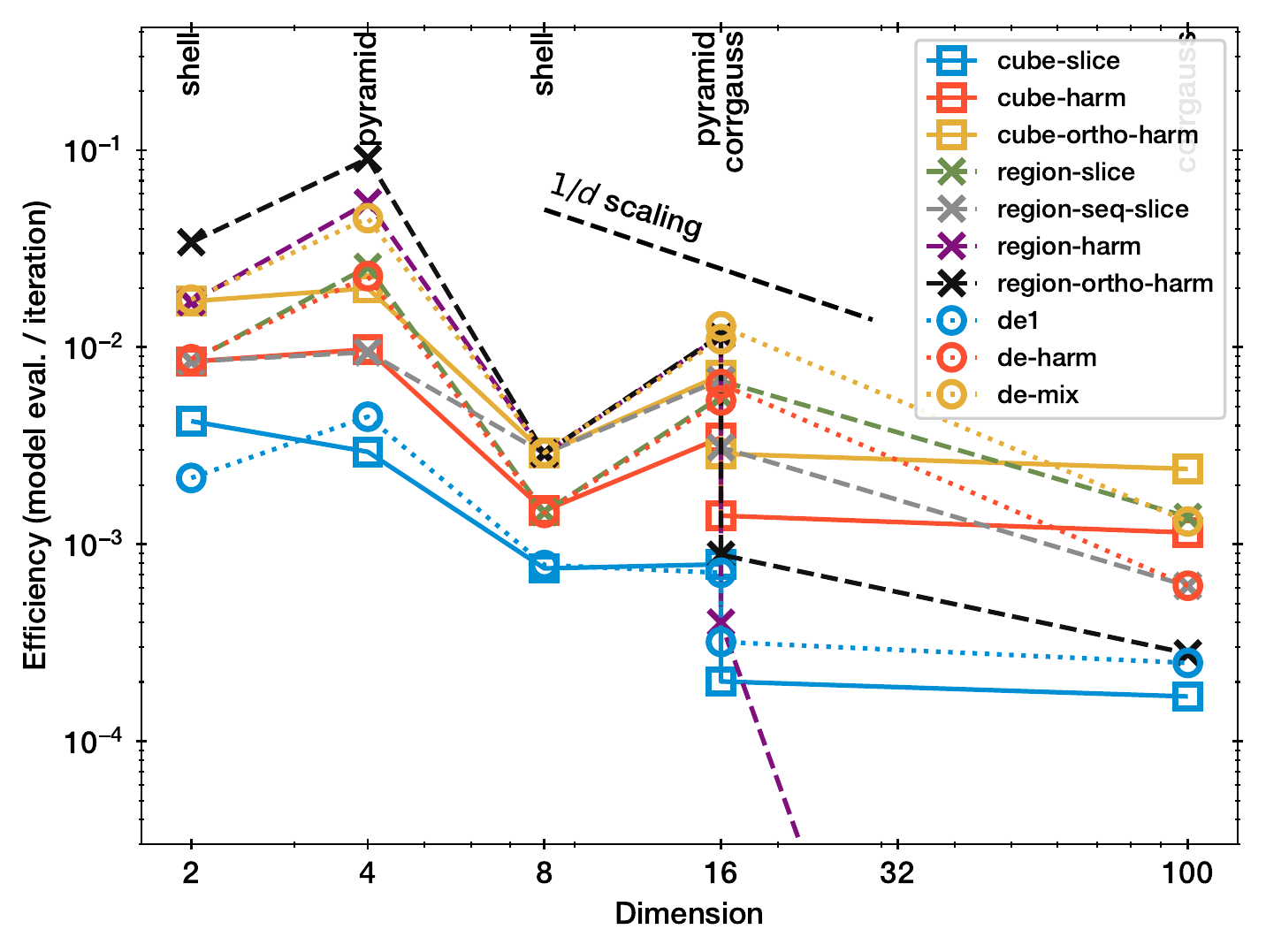}

\caption{Calibration for slice sampler.\label{sliceresult-1} and scaling.}
\end{figure}

Figure~\ref{sliceresult-1} presents the efficiency of each method.
Specifically, this is the average number of model evaluations per
NS iteration for the $N_{\mathrm{steps}}$ configuration where the
shrinkage test indicates adequate performance (Figure \ref{sliceresult-2}).
Overall, all curves in Figure~\ref{sliceresult-1} show approximately
a $1/d$ behaviour. There is substantial vertical scatter, indicating
that the efficiency is diverse across the methods and also dependant
on problem. For example, the efficiency is generally higher for the
pyramid geometry than for the more difficult shell.

\section{Discussion\label{sec:Discussion}}

Methods can now be compared by the relative position of the curves
in Figure~\ref{sliceresult-1}. Slice sampling (blue squares) is
the most inefficient among the considered methods. The single-parameter
differential proposal (blue circle) has marginally improved the efficiency.
Hit-and-run Monte Carlo (red squares) is more than twice as efficient
as these. Orthogonalisation helps further (yellow squares). The differential
proposal (de-harm) has comparable performance to these, with slightly
better behaviour at lower $d$ than at high $d$. 

The results from the whitened proposals (region-{*}) are diverse.
To inform the discussion of these results, two opposing insights seem
relevant. Firstly, whitening a space, or more accurately adjusting
the proposal to the distribution sampled is a common MCMC technique
to improve the proposal efficiency. For example, in a highly correlated
Gaussian, sampling along one axis, then another will diffuse only
short distances, while sampling along the principal axis generates
distant jumps. This may be the interpretation for the poor performance
of the cube-slice sampler (blue curve in Figure~\ref{sliceresult-1}),
relative to the region-slice (green crosses). A similar effect could
be that picking the same axis twice may lead to unnecessary reversals.
However, our results show efficiency worsens when the directions are
sampled sequentially (region-seq-slice; gray crosses) instead of randomly
(region-slice; green crosses).

Secondly, if the adaptation is computed from the live points, an iterative
bias could be induced. Indeed, it is usually recommended to perform
a warm-up or burn-in step which adjust the proposal, before a sampling
phase where the proposal is fixed and the samples are trusted. \citep{Huijser2015}
observed that in the affine-invariant ensemble sampler \citep{2010CAMCS...5...65G},
which uses differential vectors from one half of the live points to
update the other half of live points, the iterative updating can lead
to a collapse onto a sub-space and hinder convergence to the true
distribution at high dimension. Here we observe that while region-slice
performs efficiently, and cube-harm performs efficiently, region-harm
(purple crosses) performs very poorly at $d>16$. This may be a similar
self-enforcing behaviour. The orthogonalisation of region-harm (black
crosses) helps, but requires still a high number of steps at high
dimension. It is interesting that cube-harm does not show this behaviour.
From this, we can therefore conclude that hit-and-run and whitening
are good ideas separately, however the iterative estimation of covariance
matrices during the run may lead to problematic behaviour.

It is surprising that region-slice outperforms cube-slice even at
symmetric problems, such as the shell and pyramid at $d<10$. This
may be because the sample covariance estimation is an additional source
of noise and thus mixing. If this is true, then it is perhaps wiser
to not rely on this and instead use a harm step to introduce more
directional diversity, at least sometimes.

Among the algorithms discussed so far, region-slice performs best
and shows a relatively flat scaling with dimensionality. This however
may be driven by the 100d data point, by chance, although if that
data point has to be corrected down by a factor of 2, it is still
within the $1/d$ trend.

The differential evolution hit-and-run variant lies in the middle,
performing similar to the other slice and hit-and-run methods, with
no strong outliers. It outperforms cube-slice and cube-de, and behaves
similar to region-slice. This suggests that this method is not susceptible
to collapse in the same way as whitened hit-and-run.

The best performance overall is the mixture of de-harm and region-slice
(de-mix). It outperforms both individual methods. In particular, in
the shell problems, we speculate that the differential vector proposal
helps circumnavigate the shell.

Finally, how do these results apply to application to real world NS
analyses? The shrinkage test in our setup is approximately sensitive
to 1\% deviations in mean shrinkage $\left(V_{i}/V_{i+1}\right)^{K}$
over 10,000 iterations. This is directly related to the integral accuracy
in NS and corresponds approximately to the accuracy desired in real
NS runs that are dozens of times longer. Given the results in this
limited study, the de-mix, region-slice and cube-ortho-harm methods
appear promising for future research and application. The number of
steps minimally necessary for use are listed in Table~\ref{tab1}
following $N_{\mathrm{steps}}=k\times d$, together with the efficiencies
observed in this work. Future simulation studies should extend the
scope to asymmetric problems where some parameters shrinks faster
than the others and multi-modal problems, as well as to other step
samplers. The tested proposals are implemented in the \href{https://johannesbuchner.github.io/UltraNest/}{UltraNest}
nested sampling package, enabling efficient low and high-dimensional
inference of a large class of practical inference problems relevant
to cosmology, particle physics and astrophysics.

\vspace{6pt}

\supplementary{The implementation of step samplers is part of the UltraNest nested
sampling implementation available at \url{https://johannesbuchner.github.io/UltraNest/}.
Data and code for reproducing the evaluation study are available online
at \url{https://github.com/JohannesBuchner/paper-nested-sampling-stepsampler-comparison}.}

\abbreviations{Abbreviations}{The following abbreviations are used in this manuscript:\\

\noindent%
\begin{tabular}{@{}ll}
LRPS & Likelihood-restricted prior sampling\tabularnewline
NS & Nested sampling\tabularnewline
MCMC & Markov Chain Monte Carlo\tabularnewline
\end{tabular}}

\appendixtitles{no}

\appendixstart{}

\appendix

\begin{adjustwidth}{-\extralength}{0cm}{}

\reftitle{References}

\externalbibliography{yes}

\bibliography{stats}


\end{adjustwidth}{}
\end{document}